# Single-Photon Spin-Orbit Entangled States in Optical Fibers

Li Yang[1*], Ziyi Yang[1], Huaxing Xu[2]

1. Department of Electronic Engineering and Information Science, University of Science and Technology of China, Hefei 230027, Anhui, China
Key Laboratory of Electromagnetic Space Information, Chinese Academy of Sciences, Hefei 230027, Anhui, China
2. National Engineering Laboratory for Public Safety Risk Perception and Control by Big Data, China Academy of Electronics and Information Technology, Beijing 100041, China
yanglil@ustc.edu.cn*

**Abstract:** Even and odd vector modes in optical fibers are represented with linear combinations of orbital angular momentum modes, and considered as single-photon spin-orbit entangled states. It renders generating such states with fiber structures possible.

## 1. Introduction

In addition to spin angular momentum (SAM), photons can also carry orbital angular momentum (OAM), which is a new degree of freedom and a resource for quantum applications [1]. Among growing studies concerning OAM of photons, single-photon spin-orbit entanglement is of great interest, in that it can be used to test the violation of a Bell-like inequality and the invalidity of the non-contextual hidden variable (NCHV) theories [2]. Moreover, it encodes a higher-dimensional quantum space in a single photon and overcomes the limitations of the traditional information encoding based on the two-dimensional quantum space of photon polarizations (or SAM) [3]. However, optical tools for preparing such single-photon spin-orbit entangled states are limited, which are mainly based on the q plates [2,3]. Considering compatibility with existing optical fiber communication systems, one intriguing question is whether it is possible to prepare such entangled states with fiber structures. To the best of our knowledge, single-photon spin-orbit entanglement in optical fibers has not been discussed in detail, and there has been no fiber means of generating single-photon spin-orbit entanglement so far.

On the other hand, it has been proved that OAM modes can exist in optical fibers as linear combinations of conventional vector eigenmodes. Unlike the case in free-space, SAM and OAM are not independent in optical fibers. More specifically, the circular polarization direction is aligned or anti-aligned with the OAM [4]. The relationship among the angular order of vector modes and the charges of SAM, OAM and total angular momentum (AM) of the OAM modes were summarized in [5].

In this paper, we reveal the physical meaning of the angular and radial mode orders of the vector modes based on the expressions of corresponding OAM modes in optical fibers. Then, for the first time, we prove that even and odd vector modes in optical fibers can be considered as single-photon spin-orbit entangled states by representing even and odd vector modes with linear combinations of OAM modes. This theoretical analysis lays the foundation for preparing and transmitting such entangled states with optical fiber structures.

## 2. Results

### 1.1. The higher order modes with $l > 1$

In weakly guiding fibers, the transverse electric fields of higher order vector modes $HE_{l+1,m}$ and $EH_{l-1,m}$ are [6]:

$$\begin{cases} HE_{l+1,m}^{even} \\ HE_{l+1,m}^{odd} \end{cases} = F_{lm}(r) \begin{cases} \hat{x}\cos(l\phi) - \hat{y}\sin(l\phi) \\ \hat{x}\sin(l\phi) + \hat{y}\cos(l\phi) \end{cases}$$

$$\begin{cases} EH_{l-1,m}^{even} \\ EH_{l-1,m}^{odd} \end{cases} = F_{lm}(r) \begin{cases} \hat{x}\cos(l\phi) + \hat{y}\sin(l\phi) \\ \hat{x}\sin(l\phi) - \hat{y}\cos(l\phi) \end{cases} \quad (1)$$

where $l$ and $m$ stand for angular and radial mode orders, $\hat{x}$ and $\hat{y}$ are unit vectors with respect to axes $x$ and $y$, $r$ and $\phi$ are the radial and azimuthal coordinates, and $F_{lm}(r)$ denotes the radial wave functions of the scalar $LP_{lm}$ modes. One may use linear combinations of these strictly degenerated even and odd modes with a $\pi/2$ phase shift to yield OAM modes [7]:

$$\{\vec{e}_{t;lm}\} = \begin{cases} HE_{l+1,m}^{even} \pm iHE_{l+1,m}^{odd} \\ EH_{l-1,m}^{even} \pm iEH_{l-1,m}^{odd} \end{cases} = F_{lm}(r) \begin{cases} \hat{\sigma}^{\pm}\exp(\pm il\phi) \\ \hat{\sigma}^{\mp}\exp(\pm il\phi) \end{cases} \quad (2)$$

where $\hat{\sigma}^{\pm} = \hat{x} \pm i\hat{y}$ represent left and right handed circular polarization, corresponding to SAM with orders $s = \pm 1$ carried by each photon, respectively. The helical phase factors $\exp(\pm il\phi)$ correspond to OAM = $\pm l\hbar$ states. For OAM modes constituted by HE modes, the charges of the SAM and the OAM have the same sign; for EH modes, the signs of the charges are opposite. For a certain pair of mode orders ($l$, $m$), four-fold degeneracy exists, in that the polarization state and the angular distribution are both two-dimensional. In other words, the four-fold degeneracy corresponds to different combinations of SAM and OAM.

Equation (3) offers a way of understanding vector modes in optical fibers from the viewpoint of physics: the angular orders $l\pm1$ of the HE/EH modes equal to the absolute value of the charge of the total AM carried by each photon after the linear combination in (3). At the single-photon level, the four-fold degeneracy can be considered as different combinations of SAM and OAM carried by the single photon, which can be expressed as:

$$\begin{cases} |+\rangle_s |+l\rangle_l = |+,+l\rangle = F_{lm}(r)\hat{\sigma}^+ \exp(il\phi) \\ |-\rangle_s |-l\rangle_l = |-,-l\rangle = F_{lm}(r)\hat{\sigma}^- \exp(-il\phi) \\ |+\rangle_s |-l\rangle_l = |+,-l\rangle = F_{lm}(r)\hat{\sigma}^+ \exp(-il\phi) \\ |-\rangle_s |+l\rangle_l = |-,+l\rangle = F_{lm}(r)\hat{\sigma}^- \exp(il\phi) \end{cases} \quad (3)$$

where subscripts $s$ and $l$ denote SAM and OAM, respectively. The radial mode order $m$ is related to the propagation constant, which corresponds to the z-component of the linear momentum of the mode. Based on this, we can identify different modes according to their linear momentum and angular momentum.

### 1.2. The fundamental $l = 0$ modes

For fundamental $l = 0$ modes, only two-fold degeneracy exists since the angular distributions are uniform. $HE_{1m}$ modes can be expressed as:

$$\begin{Bmatrix} HE_{1,m}^{even} \\ HE_{1,m}^{odd} \end{Bmatrix} = F_{0m}(r) \begin{Bmatrix} \hat{x} \\ \hat{y} \end{Bmatrix} \quad (4)$$

combining them in a similar way as in equation (3) yields:

$$\{\vec{e}_{t;1m}\} = \{HE_{1m}^{even} \pm iHE_{1m}^{odd}\} = F_{0m}(r) \begin{Bmatrix} \hat{\sigma}^+ \\ \hat{\sigma}^- \end{Bmatrix} \quad (5)$$

where the photon carries SAM with orders $s = \pm1$ and zero OAM. This can be interpreted as:

$$\begin{cases} |+\rangle_s |0\rangle_l = |+,0\rangle = F_{0m}(r)\hat{\sigma}^+ \\ |-\rangle_s |0\rangle_l = |-,0\rangle = F_{0m}(r)\hat{\sigma}^- \end{cases} \quad (6)$$

### 1.3. The modes with $l = 1$

The $HE_{2m}^{even}$, $HE_{2m}^{odd}$, $TE_{0m}$ and $TM_{0m}$ modes are represented by the $LP_{1m}$ mode group:

$$\begin{Bmatrix} HE_{2m}^{even} \\ HE_{2m}^{odd} \end{Bmatrix} = F_{1m}(r) \begin{Bmatrix} \hat{x}\cos(\phi) - \hat{y}\sin(\phi) \\ \hat{x}\sin(\phi) + \hat{y}\cos(\phi) \end{Bmatrix}$$

$$\begin{Bmatrix} TM_{0m} \\ TE_{0m} \end{Bmatrix} = F_{1m}(r) \begin{Bmatrix} \hat{x}\cos(\phi) + \hat{y}\sin(\phi) \\ \hat{x}\sin(\phi) - \hat{y}\cos(\phi) \end{Bmatrix} \quad (7)$$

The corresponding OAM modes are:

$$\{\vec{e}_{t;1m}\} = \begin{Bmatrix} HE_{2m}^{even} \pm iHE_{2m}^{odd} \\ TM_{0m} \pm iTE_{0m} \end{Bmatrix} = F_{1m}(r) \begin{Bmatrix} \hat{\sigma}^{\pm} \exp(\pm i\phi) \\ \hat{\sigma}^{\mp} \exp(\pm i\phi) \end{Bmatrix} \quad (8)$$

where the photon carries SAM with orders $s = \pm1$ and OAM with orders $|l| = \pm1$. This can be interpreted as:

$$\begin{cases} |+\rangle_s |+1\rangle_l = |+,+\rangle = F_{1m}(r)\hat{\sigma}^+ \exp(i\phi) \\ |-\rangle_s |-1\rangle_l = |-,-\rangle = F_{1m}(r)\hat{\sigma}^- \exp(-i\phi) \\ |-\rangle_s |+1\rangle_l = |-,+\rangle = F_{1m}(r)\hat{\sigma}^- \exp(i\phi) \\ |+\rangle_s |-1\rangle_l = |+,-\rangle = F_{1m}(r)\hat{\sigma}^+ \exp(-i\phi) \end{cases} \quad (9)$$

### 1.4. Single-photon spin-orbit entanglement in optical fibers

From equation (3), (6) and (9), where OAM modes comprise even and odd vector modes, we can see that in optical fibers SAM and OAM of modes are not independent. In turn, even and odd vector modes can be represented by OAM modes:

$$\begin{cases} \left|\Phi^+\right\rangle \equiv \mathrm{HE}_{l+1,m}^{\mathrm{even}} = \frac{1}{\sqrt{2}}\left(\left|+\right\rangle_s\left|+l\right\rangle_l + \left|-\right\rangle_s\left|-l\right\rangle_l\right) \\ \left|\Phi^-\right\rangle \equiv \mathrm{HE}_{l+1,m}^{\mathrm{odd}} = \frac{1}{i\sqrt{2}}\left(\left|+\right\rangle_s\left|+l\right\rangle_l - \left|-\right\rangle_s\left|-l\right\rangle_l\right) \\ \left|\Psi^+\right\rangle \equiv \mathrm{EH}_{l-1,m}^{\mathrm{even}} = \frac{1}{\sqrt{2}}\left(\left|-\right\rangle_s\left|+l\right\rangle_l + \left|+\right\rangle_s\left|-l\right\rangle_l\right) \\ \left|\Psi^-\right\rangle \equiv \mathrm{EH}_{l-1,m}^{\mathrm{odd}} = \frac{1}{i\sqrt{2}}\left(\left|-\right\rangle_s\left|+l\right\rangle_l - \left|+\right\rangle_s\left|-l\right\rangle_l\right) \end{cases} \quad (l > 1) \qquad (10)$$

$$\begin{cases} \left|\Phi^+\right\rangle \equiv \mathrm{HE}_{2m}^{\mathrm{even}} = \frac{1}{\sqrt{2}}\left(\left|+\right\rangle_s\left|+1\right\rangle_l + \left|-\right\rangle_s\left|-1\right\rangle_l\right) \\ \left|\Phi^-\right\rangle \equiv \mathrm{HE}_{2m}^{\mathrm{odd}} = \frac{1}{i\sqrt{2}}\left(\left|+\right\rangle_s\left|+1\right\rangle_l - \left|-\right\rangle_s\left|-1\right\rangle_l\right) \end{cases} \quad (l = 1) \qquad (11)$$

where $1/\sqrt{2}$ is the normalized factor. Again, we consider the single-photon level: for modes with $l > 1$, there are four single-photon Bell states, as in equation (11). For modes with $l = 1$, the $TE_{0m}$ and $TM_{0m}$ modes are not strictly degenerate and the consequential OAM modes are unstable, so they are neglected in equation (12). As a result, there are only two single-photon Bell states in (12). Note there there is no case with $l = 0$ in the above equations since the photon carries zero OAM and there is no spin-orbit entanglement in this case.

$\left|\Phi^+\right\rangle$, $\left|\Phi^-\right\rangle$, $\left|\Psi^+\right\rangle$ and $\left|\Psi^-\right\rangle$ are each an entangled state between two qubits encoded in the SAM and OAM degrees of freedom. Equation (11) and (12) suggest that even and odd vector modes in optical fibers are spin-orbit entangled states. Thus, by generating pure even and odd modes in optical fibers, the spin-orbit entangled states can be generated.

## 3. Conclusion

Based on the expressions of OAM modes in fibers, we revealed the physical meanings of the angular and radial mode orders of the corresponding vector modes: the angular orders $l\pm 1$ of the HE/EH modes equal to the absolute value of the charge of the total AM carried by each photon after the linear combination; the radial mode order $m$ is related to the propagation constant, which corresponds to the z-component of the linear momentum of the mode. By representing even and odd vector modes with OAM modes in optical fibers, we demonstrated that the even and odd vector modes in optical fibers can be considered as single-photon spin-orbit entangled states. For modes with OAM order $l > 1$, there are four single-photon Bell states; for modes with $l = 1$, there are only two single-photon Bell states. This indicates the possibility of using optical fiber structures as a new way to generate and distribute single-photon spin-orbit entangled states.

## 4. Acknowledgments

This work was supported by the National Natural Science Foundation of China under grant 61377022 and 61705202, the Joint Funds of China Electronics Technology Group Corporation 6141B082902, and the Innovation Funds of China Electronics Technology Group Corporation.